# Sub-Cycle Optical Response Caused by Dressed State with Phase-Locked Wavefunctions

K. Uchida[1,2], T. Otobe[3], T. Mochizuki[4], C. Kim[5], M. Yoshita[5], H. Akiyama[5], L. N. Pfeiffer[6], K. W. West[6], K. Tanaka[1,2], and H. Hirori[1,7,*]

[1] *Institute for Integrated Cell-Material Sciences (WPI-iCeMS), Kyoto University, Sakyo-ku, Kyoto 606-8501, Japan*

[2] *Department of Physics, Graduate School of Science, Kyoto University, Sakyo-ku, Kyoto 606-8502, Japan*

[3] *Kansai Photon Science Institute, National Institutes for Quantum and Radiological Science and Technology, Kizugawa, Kyoto 619-0615, Japan*

[4] *Fukushima Renewable Energy Institute, National Institute of Advanced Industrial Science and Technology, Koriyama, Fukushima 963-0298, Japan*

[5] *Institute for Solid State Physics, the University of Tokyo, and JST-CREST, Kashiwa, Chiba 277-8581, Japan*

[6] *Department of Electrical Engineering, Princeton University, Princeton, New Jersey 08544, USA*

[7] *Precursory Research for Embryonic Science and Technology (PRESTO), Japan Science and Technology Agency, Kawaguchi, Saitama 332-0012, Japan*

**Abstract**

The coherent interaction of light with matter imprints the phase information of the light field on the wavefunction of the photon-dressed electronic state. Driving electric field, together with a stable phase that is associated with the optical probe pulses, enables the role of the dressed state in the optical response to be investigated. We observed optical absorption strengths modulated on a sub-cycle timescale in a GaAs quantum well in the presence of a multi-cycle terahertz driving pulse using a near-infrared probe pulse. The measurements were in good agreement with the analytical formula that accounts for the optical susceptibilities caused by the dressed state of excitons, which indicates that the output probe intensity was coherently reshaped by the excitonic sideband emissions.

PACS: 78.47.J-, 78.67.De, 42.65.Ky, 42.50.Hz

*hirori@icems.kyoto-u.ac.jp2

The ability to exploit coherent light-matter interactions lies at the heart of controlling the dynamical behavior of electronic systems for the purpose of developing ultrafast switching devices [1], quantum information processing devices [2], and attosecond laser technology [3]. In general, electronic systems under laser driving fields form a photon-dressed state, leading to quantum optical phenomena, such as the optical Stark effect (Autler-Townes splitting) [4-7], electromagnetically induced transparency [8], and gain without population inversion [9,10]. The dominant physics underlying the photon dressing involves strong coupling between the electronic system and laser field through the coherent interaction. When the driving field of $E(t)$ has temporal periodicity $T = 2\pi/\Omega$, i.e., $E(t) = E\cos(\Omega(t+\tau))$, Floquet theory can be used to solve the time-dependent Schrödinger equation of the dressed state $\psi_e$, as follows [11,12]:

$$\psi_e(t) = \exp(-i\varepsilon_e t/\hbar)\sum_l \psi_{e,l}\exp[-il\Omega(t+\tau)], \qquad (1)$$

where $\varepsilon_e$ is the quasienergy, $\psi_{e,l}$ ($|e,l\rangle$) is the $l$th sideband state of the dressed state $\psi_e$, $\tau$ is the time delay between the driving field $E(t)$ and the instantaneous time of the observation, and $\hbar$ denotes Planck's constant divided by $2\pi$.

The essential aspect of the photon dressing is imprinting the phase information of the driving field on the quantum phase of the dressed state through $l\Omega\tau$ in Eq. (1), as well as the energy spacing of the sidebands in units of the driving photon energy $\hbar\Omega$, as shown in Fig. 1(a). When the dressed state $\psi_e$ is excited by the optical field of the probe pulse at a time delay $\tau = \tau_1$, even considering the case of an optical component exciting $|e,l\rangle$ ($l = 0$) (Fig. 1(b)), because the $\psi_e$ is a quantum superposition state described by Eq. (1) the polarizations between the vacuum $|0\rangle$ and the every different sideband states $|e,l'\rangle$ can be induced and leads to the phase-locked sideband emissions [13,14]. Their phase

difference depends on the indices *l* and *l'* of the sideband according to $(l'-l)\Omega\tau$ and sideband emissions can dynamically change the output intensity spectra with $\tau$ and $\Omega$, thereby providing a means of developing novel optical synthesizing and processing devices in which the quantum phases of the dressed state carry information [15]. To date, although many fascinating optical phenomena featuring the dressed states have been studied in a variety of materials [13,14,16,17], the studies are restricted to the driving-cycle-averaged experiments and thus our understanding of the role of quantum phases of the dressed state in the optical response remains limited.

In this work, we show the sub-cycle modulation of optical absorption near exciton state in GaAs quantum wells (QW) under the presence of carrier envelope phase (CEP)-stable and strong terahertz (THz) transients (Fig. 1(c)). The unique merit of a THz pulse generated using a femtosecond laser is that the waveforms have an inherently stable CEP [18-21], which allows the optical response to be measured in a phase-sensitive manner [22,23]. The frequency analysis of the temporal responses shows the formation of the excitonic dressed state that is a quantum superposition state with energetically spaced sidebands. The measurements were in good agreement with the analytical formula that accounts for the optical susceptibilities caused by the excitonic dressed state, indicating that the output probe intensity was coherently reshaped by the excitonic sideband emissions.

The sample consisted of GaAs QWs with widths of 12 nm that are separated by 10-nm $Al_{0.3}Ga_{0.7}As$ barriers (ten periods) grown by molecular beam epitaxy, and the QW were sandwiched between 1-μm $Al_{0.33}Ga_{0.67}As$ digital-alloy-barrier layers. For optical transmission measurements, the GaAs substrate was removed by chemical etching, and then the sample was glued to an $SiO_2$ substrate. The laser source was a Ti:sapphire regenerative amplifier (repetition rate: 1 kHz, pulse width: 100 fs, central photon energy:

1.55 eV, pulse energy: 4 mJ/pulse). The pump THz pulses were generated by optical rectification in a LiNbO$_3$ crystal with the tilted-pulse-intensity-front scheme [19,21]. The temporal profiles of the THz pulses at the sample position were measured using electro-optic (EO) sampling with a GaP crystal attached to the same SiO$_2$ substrate. A bandpass filter was used to convert single-cycle THz pulses into linearly polarized multi-cycle ones that could be regarded as a temporally periodic field. All of the experiments were performed at 10 K, and the sample was designed such that the NIR probe spectra with bandwidth of 17 meV [full width at half maximum (FWHM) intensity] would cover the excitonic resonance absorption. The transmitted probe pulse was spectrally resolved using a spectrometer and then detected using a charge-coupled device (CCD) [24].

Figures 2(a) and 2(b) show that the driving field of the pump THz pulse has a fairly sinusoidal temporal shape in the sample and a central frequency (photon energy) of $\Omega/2\pi$ = 0.6 THz ($\hbar\Omega$ = 2.5 meV). Figure 2(c) shows the absorption $\alpha$ of the sample without the THz pump (shaded gray area), exhibiting the salient absorption of the 1s heavy-hole exciton (ex1) at $\varepsilon_{ex1}$ =1.55 eV, which is below the bandgap energy of 1.56 eV. The intraexcitonic transition energies with large dipole moments lie in the THz frequency range [26,27] making excitons useful systems not only for studying nonlinear optics but also for exploiting THz excitonic interactions in solid-based devices [28-30]. The intra-excitonic 1s-2p transition energy $\hbar\Delta$ is estimated to be ~8.2 meV, leading to the non-resonant THz interaction with the transition as shown in Fig. 2(d). The time-averaged spectrum (orange dotted line in Fig. 2(c)) with THz field $E$ = 0.5 kV/cm over $\tau$ between −1.5 and 2.5 ps does not exhibit any remarkable structures. Nonetheless, the time-resolved absorption changes present remarkable dynamical features: the left panel of Fig. 2(e) shows that the absorption changes $\Delta\alpha$ sliced at $\varepsilon_{ex1}+2\hbar\Omega$ (thick blue solid line) and at $\varepsilon_{ex1}$ (red solid line) in the right panel are modulated on timescales two times faster than the THz cycle period.

The narrow bandwidth of the THz pulse (~100 GHz) enables an excitonic dressed state $\psi_{ex}$ to be induced, and the subsequent probe pulse excitation causes sideband emissions from it (Fig. 1(b)). The optically allowed sidebands have an energy spacing that consists of even multiples of the THz photon energy, i.e., $2m\hbar\Omega$, where $2m = |l'-l|$ because of the spatial inversion symmetry of the system. Because the sidebands are superposed and their phases are locked to each other, the relative phases between the emissions and probe pulse develop synchronously from 0 to $2\pi$ with $\tau$, such that constructive and destructive interferences alternately appear with a differential frequency (periodicity) of $2m\Omega$ ($T/2m$).

Assuming the interaction of THz pulse with the 1s and 2p excitonic transition as shown in Fig. 2(d) [31], the optical absorption $\alpha(\omega,\tau)$, which is defined as $\alpha(\omega,\tau) \equiv \omega/n_0 c \Im[P(\omega,\tau)/\varepsilon_0 f(\omega)]$ (where $P(\omega,\tau)$ is the polarization induced by the probe pulse excitation with a spectrum $f(\omega)$, $\varepsilon_0$ is the vacuum permittivity, $n_0$ is the real part of the refractive index, and $c$ is the speed of light), can be calculated by using a standard first-order perturbation approximation [32-34]:

$$\alpha(\omega,\tau) = \frac{\omega}{n_0 c} \sum_{n,l,l'} \frac{f(\omega-(l'-l)\Omega)}{f(\omega)} \left( \Im[\chi_{n,l,l'}(\omega)]\cos(l'-l)\Omega\tau + \Re[\chi_{n,l,l'}(\omega)]\sin(l'-l)\Omega\tau \right), \quad (2)$$

$$\chi_{n,l,l'}(\omega) = \frac{|P_{CV}|^2 \phi_{n,l'}^*(r=0)\phi_{n,l}(r=0)}{\varepsilon_0 \hbar(\omega - \varepsilon_n/\hbar - l'\Omega - i\Gamma)}, \quad (3)$$

where $P_{CV}$ is the interband dipole moment, $\varepsilon_n$ is the excionic quasienergy [38], $\Gamma$ is the phenomenological damping, $\Re$ and $\Im$ respectively denotes the real and imaginary part, $\phi_{n,l}$ indicates the envelope function of relative motion of $n$th exciton whose sideband index is $l$ [21], and $r$ is the relative position of electron and hole pair. $\chi_{n,l,l'}(\omega)$ indicates the optical susceptibility of the dressed states for $n$th exciton states, $n = $ ex1 and ex2 for the 1s and 2p excitonic states in our calculation, and represent the optical process wherein the excitation of the *l*th sideband induces emission from the *l*'th sideband.

The calculation results presented in Fig. 2(e) demonstrate that our model captures the observed dynamical behaviors very well (i.e., periodic changes on timescales two times faster than the THz cycle period at both $\varepsilon_{ex1} + 2\hbar\Omega$ and $\varepsilon_{ex1}$.) In addition, the analytical formula $\alpha(\omega,\tau)$ (Eq. (2)), which includes the alternation between the real and imaginary parts of the susceptibility $\chi_{n,l,l'}(\omega)$ depending on $\tau$, reproduces the tendency that the phase lag becomes large with the distance from $\varepsilon_{ex1} + 2\hbar\Omega$ (the dotted lines in Figs. 2(e) and 2(f)). These results imply that the nonlinear susceptibility that defines the output optical intensity can be controlled by changing the timing of the optical probe excitation.

The increase in the THz field can be expected to induce faster variations than the half-cycle periodicity with a time delay $\tau$ because it increases the THz excitonic interaction strength as $dE$ (= $\Omega_R$: Rabi energy), where $d$ is the dipole moment of the intra-excitonic transition, and induces higher sidebands ($|l| > 2$), which may contribute to the optical response. To observe the effects of higher sidebands, $\Delta\alpha$ was measured with a relatively stronger field ($E = 1.6$ kV/cm), as shown in Fig. 3(a). Figure 3(b) shows temporal profiles of $\Delta\alpha$ sliced at $\varepsilon_{ex1}+2\hbar\Omega$ and $\varepsilon_{ex1}+4\hbar\Omega$, which reveal oscillations with a quarter-cycle $T/4$ (green line) and half-cycle $T/2$ periodicity (orange line). The Fourier transform spectrum of $\Delta\alpha$ shown in Fig. 3(a) reveals the origins of the observed sub-cycle responses (Fig. 3(c)). The solid lines in Figs. 3(d-f) show the experimental 2*m*th (2*m* = |*l*−*l*'| = 2, 4 and 6) harmonic spectra; there is a peak at photon energies of $\varepsilon_{ex1}+2m\hbar\Omega$ aside from the one at $\varepsilon_{ex1}$. The peak energy increases as an increase of $\hbar\Omega$ [39], and these spectral features can be reproduced by calculations with a perturbation approximation extending to 2*m*th order (dashed lines), allowing us to attribute the peaks to the sideband emissions at $\varepsilon_{ex1}+2m\hbar\Omega$ as origin of the sub-cycle response [40].

The phase-resolved measurement of the nonlinear optical response clarifies the role of the nonlinear susceptibility in the sub-cycle absorption changes. Figure 4(a) shows a schematic diagram of the optical process inducing the nonlinear susceptibility $\chi_{ex1,4,0}$, which dominates the fourth harmonics of $\Delta\alpha$ near $\varepsilon_{ex1}$. Figures 4(b) and 4(c) show the fourth harmonics spectra of the $\Delta\alpha$ at time delays of $\tau = 0$ and $\tau = T/16$ extracted by reconstructing the Fourier amplitude and phase spectra obtained experimentally and the corresponding ones calculated using Eqs. (2) and (3). These spectra exhibit similar Lorentzian-type spectral shapes, reflecting the $\chi_{ex1,4,0}$ described by Eq. (3). These results demonstrate that, when optically probing the quantum superposition state, the real and imaginary part of $\chi_{n,l,l'}(\omega)$ alternately appears in the $\Delta\alpha$ with a beating frequency $(l'-l)\Omega$, i.e., $(0-4)\Omega = -4\Omega$ in this case (as shown in the diagram on the right side of Fig. 4(a)), thus leading to the quarter-cycle modification.

Figures 4(d-e) show the field dependence of the $\Delta\alpha$ amplitude of the $2m$th harmonics at $\varepsilon_{ex1} + 2m\hbar\Omega$. In the weak-field region ($E < \sim 2$ kV/cm), the amplitudes follow a $2m$th power law of the electric field ($\propto E^{2m}$), as indicated by the dashed lines. These results indicate that the THz fields interact with the excitonic system in a perturbative manner; thus, as shown in Fig. 2(c), the excitonic dressed state hardly differs from the bare exciton state. As the THz field increases ($E > 2$ kV/cm), the field dependences start to saturate, indicating a breakdown of perturbation theory (the Keldysh parameter $KP$ is much less than unity; $KP = \sqrt{E_B/2U_p}$ [14,41], where $E_B$ is the Coulomb binding energy and $U_p = e^2E^2/4m^*\Omega^2$ is the ponderomotive energy ($e$: elementary charge, $E$: field amplitude, $m^*$(=0.054 $m_0$): reduced mass, and $m_0$: free-electron mass)). In the excitonic system that we studied, $KP$ becomes unity at $E_{THz} \sim 3$ kV/cm because $E_B$ is $\sim 8$ meV; thus, the system exhibits saturation behavior. As $E$ increases, the phase lag becomes small; this phenomenon indicates the emergence of a non-perturbative (quasi-static) interaction [42].

In summary, the ability to produce CEP-stable strong THz transients has enabled a new field of study of coherent light-matter interactions in solids. The findings presented here go beyond the observation of the ladder-like energy levels caused by formation of the excitonic dressed state; in particular, we demonstrated that CEP-stable THz transients enable the phase of wavefunctions to be imprinted with the phase information of the driving field, and this ability may lead to precise control of the nonlinear susceptibility that causes sub-cycle optical responses. Besides showing the feasibility of this novel concept for sub-cycle control of optical properties, our findings also provide a benchmark for the interpretation of phase resolved experiments under intense light fields with a broad frequency range from terahertz to petahertz over a broad range of exotic materials from atomic layer semiconductors to strongly correlated electron systems [22,23].

strongly compared with the other higher excitonic and continuum states. A more detailed analysis considering the higher states and the many-body Coulomb interactions between carriers is beyond the scope of this report.

**Acknowledgements:**

This study was supported by KAKENHI (26286061) and partly by KAKENHI (26247052) from JSPS. The work at Princeton University was also funded by the Gordon and Betty Moore Foundation through the EPiQS initiative Grant GBMF4420, and by the National Science Foundation MRSEC Grant DMR-1420541.

**Figure captions:**

**Figure 1.** Optical probing scheme of the photon-dressed state. Energy level diagram of (**a**) the bare state ($|0\rangle$ and $|e\rangle$) and (**b**) the dressed state formed under a driving field. For the sake of clarity, the higher dressed manifolds are not shown. The wavy blue lines at each energy level represent the phase difference of the wavefunctions for the sidebands depending on the time delay $\tau$. Excitation of the dressed state by a NIR probe pulse (red solid arrow) at the time delay $\tau = \tau_1$ causes emission lines (orange solid arrows). (**c**), Schematic diagram of the experimental setup for the THz-pump and NIR-probe measurement on GaAs QWs.

**Figure 2.** Sub-cycle optical response near the bandgap energy. (**a**), Power spectrum of the pump THz pulse. (**b**), Temporal profile of the pump THz electric field measured via electro-optic sampling. (**c**), Measured absorption spectra $\alpha$ of the GaAs QW sample with the THz pump ($E = 0.5$ kV/cm) and without it (orange dotted line and the gray filled area). The label $\varepsilon_{ex1}$ indicates the peak position of the 1s heavy-hole exciton. The spectrum with the pump (orange dotted line) was time-averaged over the twice period of the THz transient ($4\pi/\Omega = 3.3$ ps). (**d**), Energy diagram of the exciton state. The THz photon and 1s-2p transition energies are $\hbar\Omega=2.5$ meV and $\hbar\Delta=8.2$ meV. (**e**), Left panel: change in the absorption $\Delta\alpha(\tau)$ as a function of the time delay $\tau$ sliced at $\varepsilon_{ex1}$ (red solid line) and $\varepsilon_{ex1}+2\hbar\Omega$ (blue solid line), as indicated by the dashed lines in $\alpha(\omega,\tau)$ (the right panel). Here, the absorption change is defined as $\Delta\alpha = -\log(I_{on}/I_{off})$, where $I_{on}$ and $I_{off}$ are the intensities of transmitted light with and without the pump pulse, respectively. (**f**), Corresponding calculated change in the absorption spectrum according to perturbation theory. The dotted lines are guides for the eye.

**Figure 3.** Effects of higher sidebands on the optical responses. (**a**), Change in the absorption spectrum $\Delta\alpha(\omega,\tau)$ at $E = 1.6$ kV/cm. The dashed lines indicate the energies of $\varepsilon_{ex1}$, $\varepsilon_{ex1}+2\hbar\Omega$, and $\varepsilon_{ex1}+4\hbar\Omega$. (**b**), Temporal profiles of $\Delta\alpha$ sliced at $\varepsilon_{ex1}+2\hbar\Omega$ (orange line) and $\varepsilon_{ex1}+4\hbar\Omega$ (green line). (**c**), The Fourier amplitude spectrum $|\Delta\alpha|$ as a function of probe photon energy $\hbar\omega$ at $E = 1.6$ kV/cm. The dashed lines are guides for the eye. (**d-f**), Fourier amplitude spectra of the second ($2m = 2$), fourth ($2m = 4$), and sixth ($2m = 6$) harmonics. The solid lines indicate the experimental results, and the dashed lines represent the calculated ones, which are scaled to the experiment. The dotted lines are guides for the eye.

**Figure 4.** Role of the nonlinear susceptibility in the sub-cycle response and field dependence of sidebands. (**a**), Schematic diagram of the sideband emission that induces the optical response. (**b**), $\Delta\alpha$ spectra of the fourth harmonic component for time delays of $\tau = 0$ (black solid line) and $\tau = T/16$ (blue dashed line), extracted by reconstructing the Fourier amplitude and phase spectra obtained experimentally. (**c**), The corresponding calculated absorption change spectra of the fourth harmonic component $\chi_{ex1,0,4}$. (**d-f**), THz field dependences of the harmonic amplitudes at $\varepsilon_{ex1}+2\hbar\Omega$ ($2m = 2$), $\varepsilon_{ex1}+4\hbar\Omega$ ($2m = 4$), and $\varepsilon_{ex1}+6\hbar\Omega$ ($2m = 6$) shown in Figs. 3(**d-f**). The dashed lines are guides for the eye that are proportional to $E^{2m}$.

# Figure 1.

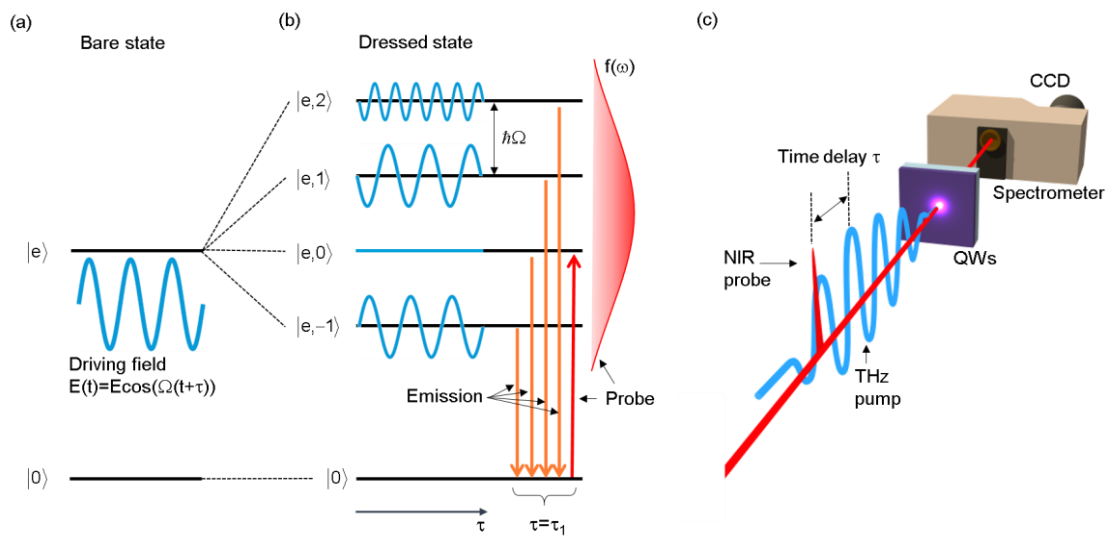

# Figure 2.

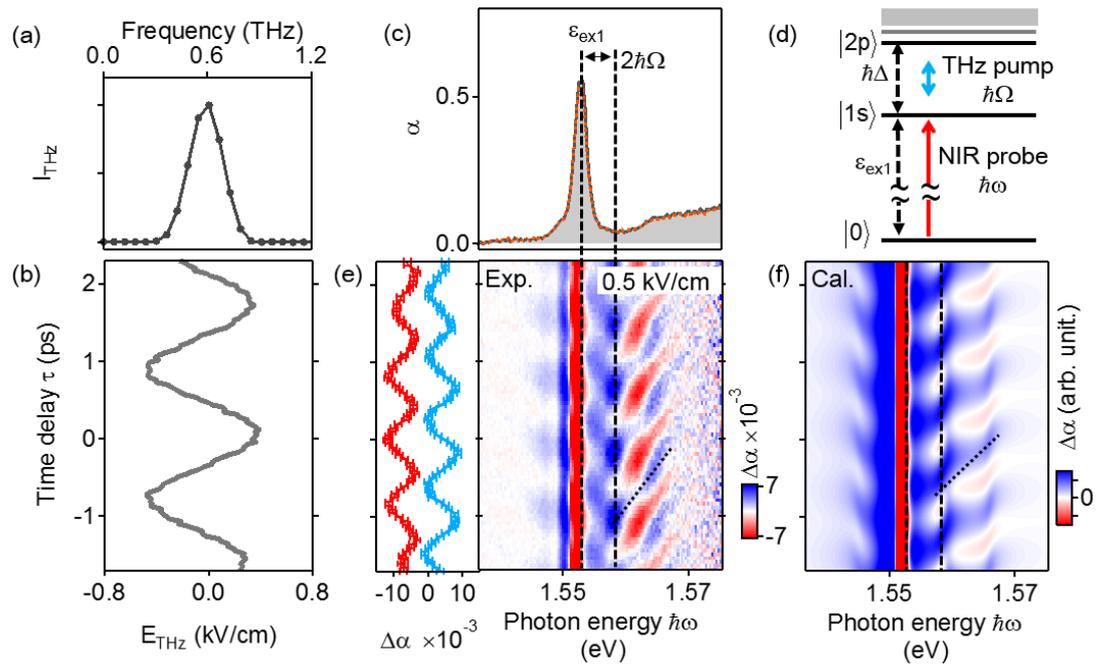

# Figure 3.

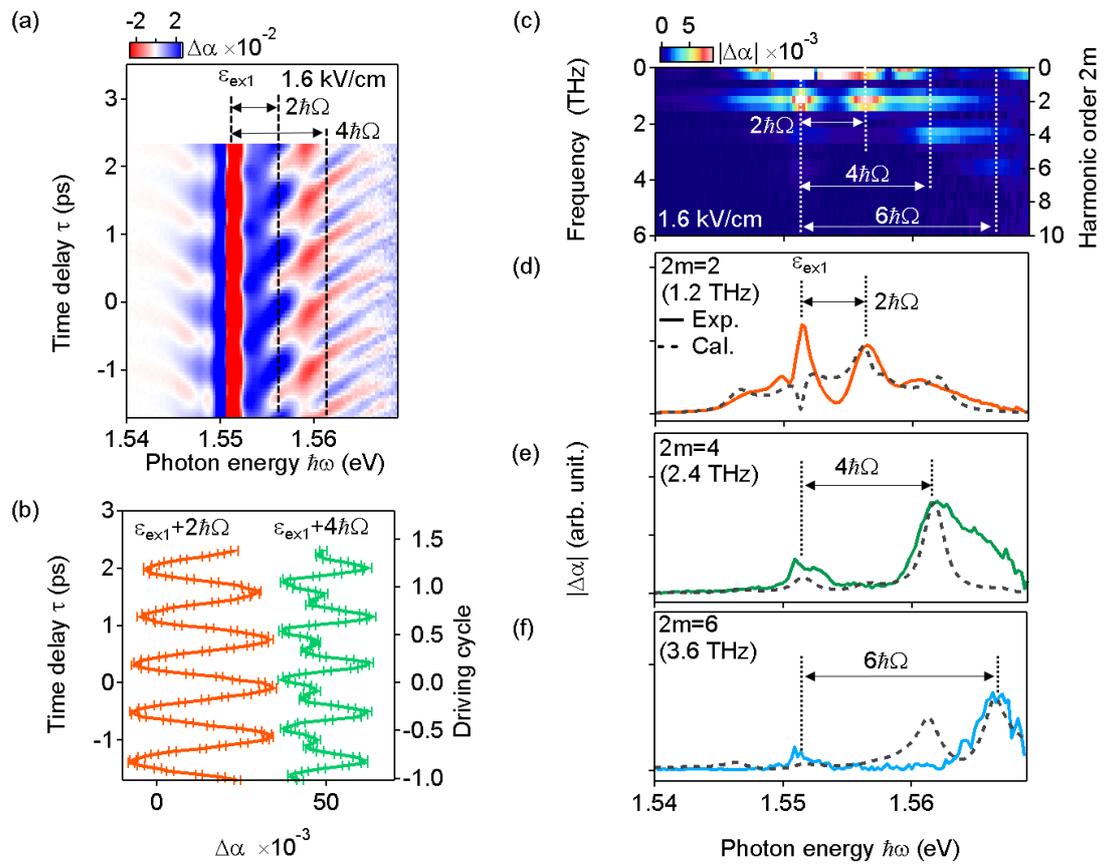

# Figure 4.

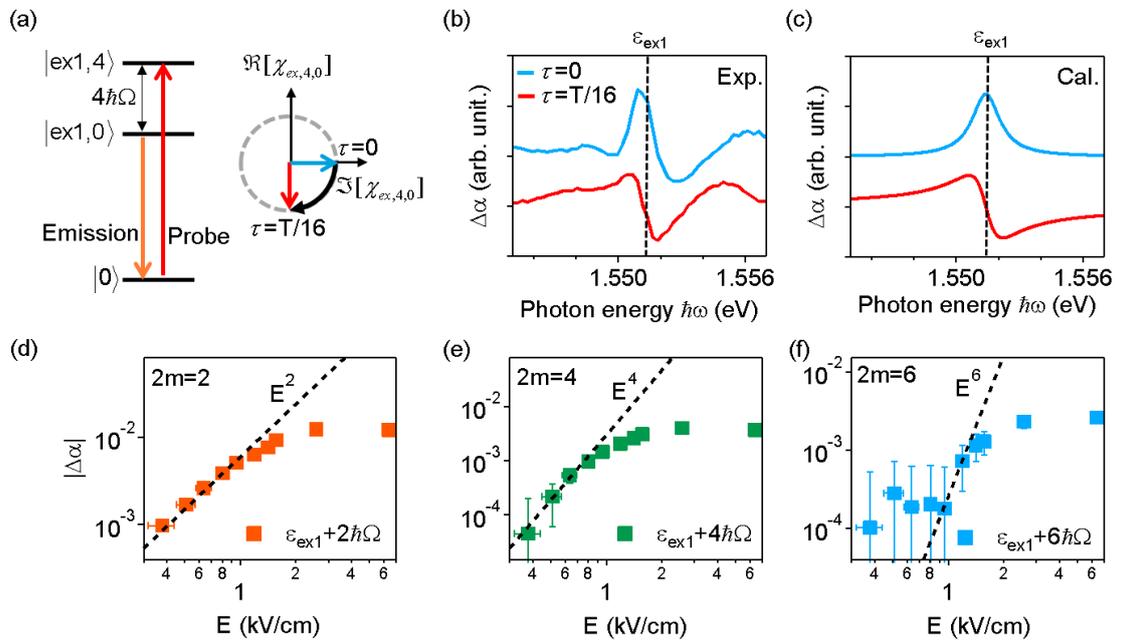